\documentclass[epj]{webofc}
\usepackage[utf8]{inputenc}
\usepackage[varg]{txfonts}   
\usepackage{booktabs}
\usepackage{xcolor}
\definecolor{darkred}{rgb}{0.4,0.0,0.0}
\definecolor{darkgreen}{rgb}{0.0,0.4,0.0}
\definecolor{darkblue}{rgb}{0.0,0.0,0.4}
\usepackage[bookmarks,linktocpage,colorlinks,
    linkcolor = darkred,
    urlcolor  = darkblue,
    citecolor = darkgreen]{hyperref}

\usepackage{bbold}

\wocname{EPJ Web of Conferences}
\woctitle{Lattice2017}
\newcommand{\SUN}{\mathrm{SU}(N)}
\newcommand{\SUNN}{\mathrm{SU}(N-1)}
\newcommand{\U}{\mathrm{U}(1)}
\newcommand{\SUtwo}{\mathrm{SU}(2)}

\begin{document}
%
\selectlanguage{english}
\title{%
A study of how the particle spectra of SU($N$) gauge theories with a fundamental Higgs emerge
}
\author{%
\firstname{Pascal} \lastname{T\"orek}\inst{1}\fnsep\thanks{Speaker, \email{pascal.toerek@uni-graz.at}}\and
\firstname{Axel} \lastname{Maas}\inst{1} \and
\firstname{Ren\'e}  \lastname{Sondenheimer}\inst{2}
}
\institute{%
Institute of Physics, NAWI Graz, University of Graz, Universit\"atsplatz 5, 8010 Graz, Austria
\and
Theoretisch-Physikalisches Institut, Friedrich-Schiller-Universit\"at Jena, Max-Wien-Platz 1, 07743 Jena, Germany
}
\abstract{%
  In gauge theories, the physical, experimentally observable spectrum consists only of gauge-invariant states. 
  In the standard model the Fr\"ohlich-Morchio-Strocchi mechanism shows that these states can be 
  adequately mapped to the gauge-dependent elementary W, Z, Higgs, and fermions. 
  In theories with a more general gauge group and Higgs sector, appearing in various extensions of the 
 standard model, this has not to be the case. In this work we determine analytically the  physical spectrum of  
 $\mathrm{SU}(N>2)$ gauge theories with a Higgs field in the fundamental representation. We show that 
 discrepancies  between the spectrum predicted by perturbation theory and the observable physical spectrum arise. 
 We confirm these analytic findings with lattice simulations for $N=3$. 
}
\maketitle

\section{Introduction}\label{intro}
The physical states of gauge theories are gauge invariant. In QCD confinement takes care of this, but in 
the electroweak sector of the standard model this is far more subtle \cite{'tHooft:1979bj,Frohlich:1980gj}: The W/
Z bosons, the Higgs, and the fermions, i.e. the elementary fields of the Lagrangian, are not gauge-invariant. Thus, 
they should not be observable. Nonetheless, using them as if they were physical in a perturbative description within 
a fixed gauge describes experimental results remarkably well \cite{pdg}.

This apparent contradiction is resolved by the Fr\"ohlich-Morchio-Strocchi (FMS) mechanism 
\cite{Frohlich:1980gj,Frohlich:1981yi}: Under certain conditions, fulfilled by the standard model, the properties of 
the physical states can be mapped to the gauge-dependent states which appear in the Lagrangian. This mechanism 
has been confirmed in lattice computations for the bosonic sector \cite{Maas:2012tj,Maas:2013aia}.

However, the one-to-one correspondence between gauge-invariant and gauge-variant quantities relies on the 
special structure of the standard model, where the gauge group and the global (custodial) symmetry group are the 
same, SU(2). 
It is therefore not guaranteed to work also in beyond the standard model (BSM) scenarios \cite{Maas:2015gma}. 
Here, we investigate the consequences for BSM-like structures as they appear, for instance, in 
grand unified theories (GUTs). Especially, we study SU($N>2$) gauge theories with a single Higgs field in the 
fundamental representation of the gauge group. 
In this case the gauge group is much larger than the global symmetry group, which is $\U$. After analytically 
determining the spectrum using the FMS mechanism in the general case, we test, and confirm, the predictions using 
lattice simulations for the special case $N=3$. In particular, we focus on states with quantum numbers not 
sustainable by the elementary particles.

We refer to \cite{Torek:2016ede}, for a brief review of the FMS prescription. For results on other theories 
arising typically in BSM scenarios see \cite{Maas:2017xzh}.

\section{SU(N>2) gauge theories with a Higgs in the fundamental representation}\label{sec-1}
We study an $\mathrm{SU}(N>2)$ gauge theory with one fundamental scalar field\footnote{For the $\SUtwo$ 
case see \cite{Frohlich:1980gj,Frohlich:1981yi,Maas:2012tj,Maas:2013aia,Maas:2016qpu}.} with a Lagrangian
\begin{align}
\mathcal{L} &= -\frac{1}{4}F_{\mu\nu}^aF^{a~\mu\nu} + (D_\mu \phi)^\dagger (D^\mu\phi) - 
V(\phi^\dagger\phi)\;,
\label{eq:lagrangian}
\end{align}
where the gauge fields $A_\mu$ with field-strength tensor $F_{\mu\nu}$ couple through the covariant derivative 
$D_\mu$ to the Higgs field $\phi$. The Higgs field is a complex $N$-component vector transforming under gauge 
transformations as $\phi(x)\rightarrow U(x)\phi(x)$ with $U(x)=e^{i \alpha^a(x) T^a}$ and $T^a$ are the 
generators of the associated Lie algebra and $\alpha^a$ arbitrary real functions. 
The Higgs potential $V$ depends on the gauge-invariant quantity $\phi^\dagger \phi$. We further allow only 
for potentials which consist of operators that are renormalizable by power counting and which have classically one 
or more minima at non-zero Higgs field. The theory has an additional global symmetry, a $\U$ custodial 
symmetry acting only on the Higgs field.

\subsection{Gauge-variant description in a fixed gauge}\label{sec-2}
To investigate the mass spectrum of the elementary fields in a fixed gauge with a nonvanishing vev for the Higgs 
field, we split the scalar field into its vev $v$ and a fluctuation part $\varphi$ around the vev, i.e.
\begin{align}
\phi(x) &= \frac{v}{\sqrt{2}}n + \varphi(x)\;,
\label{eq:phivev}
\end{align}
where $n$ is a unit vector in gauge space, $n^\dagger n = 1$, pointing in the direction of the vev. Without loss of 
generality we will usually set $n_i = \delta_{i,N}$, since we can always perform a gauge transformation such that 
the vev is in the real part of the $N^{\text{th}}$ component at every space-time point\footnote{We ignore the 
possible effect of gauge defects.}. The Higgs boson and the Goldstone bosons can be described in a 
gauge-covariant (but not gauge-invariant) manner without specifying $n$ by 
$h=\sqrt{2}\text{Re}[n^\dagger\phi]$ and $\hat{\varphi} = \phi-\text{Re}[n^\dagger\phi]n$. 

Here we implement the $R_\xi$ gauge condition
\begin{align}
\mathcal{L}_{gf} &= -\frac{1}{2\xi}\left|\partial^\mu A_\mu^a 
+ \frac{igv\xi}{\sqrt{2}}\Big(n^\dagger T^a \varphi - \varphi^\dagger T^a n\Big) \right|^2\;,
\label{eq:rxi}
\end{align}
in order to render the functional integral well-defined. The vev minimizes the potential and thus 
$\partial_\phi V|_{\phi=\frac{v}{\sqrt{2}}n}=0$. Due to the condition \eqref{eq:rxi} a mass term for the would-be 
Goldstone bosons is introduced which is proportional to the gauge-fixing parameter $\xi$. Since the Goldstone 
bosons form a BRST quartet with the ghost sector and time-like gauge bosons they will not appear in the spectrum, 
and in particular drop out in any vacuum correlator. Therefore, they will play no role in the following. 

Rewriting the scalar kinetic term in the Lagrangian \eqref{eq:lagrangian} with help of Eq. \eqref{eq:phivev}, we 
obtain
\begin{align}
(D_\mu\phi)^\dagger (D^\mu \phi) &= \partial_\mu \varphi^\dagger \partial^\mu \varphi + 
\frac{g^2v^2}{4}~A_\mu ^a ~n^\dagger\{T^a,T^b\}\,n~ A^{b~\mu} + 
\sqrt{2}gv~\text{Im}[n^\dagger T^a \partial^\mu\varphi]~A_\mu^a + \cdots\;,
\end{align}
where the usual mass matrix \cite{Bohm:2001yx} for the gauge bosons appears in the second term and the mixing 
between the longitudinal parts of the gauge bosons and the Goldstone bosons is visible in the third term, which is 
removed by the gauge condition \eqref{eq:rxi}. The neglected parts involve the interactions between three 
and four fields. 

For our choice $n_i=\delta_{i,N}$, the mass matrix $(M^2_A)^{ab}$ of the gauge bosons is given by, 
\begin{align}
\Big(M^2_A\Big)^{ab} &= \frac{g^2v^2}{2}~n^\dagger\{T^a,T^b\}\, n = 
\frac{g^2v^2}{4}~\text{diag}\Bigg(\underbrace{0,\cdots,0}_{(N-1)^2-1},
\underbrace{1,\cdots,1}_{2(N-1)},\frac{2}{N}(N-1)\Bigg)^{ab}\;.
\label{eq:gaugemass}
\end{align}
Therefore, we obtain $(N-1)^2-1$ massless gauge bosons, $2(N-1)$ degenerate massive gauge bosons with mass 
$m_A=\frac{1}{2}gv$, and one gauge boson with mass $M_A=\sqrt{2(N-1)/N}m_A$. 
Additionally, from the quadratic terms in $\varphi$ of the potential $V$ one obtains a mass $m_h=\lambda v$ for 
the elementary Higgs field, where $\lambda$ is the four-Higgs coupling, 
$\frac{\lambda}{2}(\phi^\dagger\phi)^2$.

The situation is now that which, in an abuse of language, is usually called 'spontaneously broken'. The breaking 
pattern is $\SUN\rightarrow \SUNN$. With respect to the subgroup $\SUNN$ the gauge bosons are in the adjoint 
representation (massless), a fundamental and an anti-fundamental representation (mass $m_A$) and a singlet 
representation (mass $M_A$), explaining their degeneracy pattern.

\subsection{Gauge-invariant spectrum}\label{sec-3}

In this subsection we discuss the gauge-invariant, and thus experimentally observable, spectrum of the theory 
described by Eq. \eqref{eq:lagrangian}. These states are either singlets or non-singlets with respect to the $\U$ 
global custodial symmetry. In a second step we predict their masses using the FMS mechanism.

Let us start with the singlet states \cite{Maas:2016ngo}: A gauge-invariant composite operator describing a scalar, 
positive parity boson, i.e. $J^P_{\U} = 0^+_0$, is $O_{0^+_0}(x) = (\phi^\dagger\phi)(x)$. We apply the FMS 
mechanism and expand the ensuing correlation functions to leading order, yielding \cite{Torek:2016ede}
\begin{align}
\Big\langle O_{0^+_0}(x) O^\dagger_{0^+_0}(y)\Big\rangle = \frac{v^4}{4} + 
v^2\big\langle h(x)h(y)\big\rangle_{\text{tl}} +  \big\langle h(x)h(y)\big\rangle_{\text{tl}}^2 + 
\cdots\;,
\label{eq:fmshiggs}
\end{align}
where 'tl' means 'tree level'. The second term on the r.h.s.\ of Eq. \eqref{eq:fmshiggs} describes the propagation 
of a single elementary Higgs boson and the third term describes two non-interacting Higgs bosons propagating 
both from $x$ to $y$. Comparing poles on both sides of Eq. \eqref{eq:fmshiggs} predicts the mass of the left-
hand side, and thus of the observable particle. This scalar boson should therefore have a mass equal to the mass of 
the elementary Higgs. Also, a next state should exist in this quantum number channel which is a scattering state of 
twice this mass.

Next, consider a singlet vector operator $O^\mu_{1^-_0}(x)=i(\phi^\dagger D^\mu\phi)(x)$. The same expansion 
yields
\begin{align}
\Big\langle O^\mu_{1^-_0}(x) O^{\nu~\dagger}_{1^-_0}(y) \Big\rangle &= 
\frac{(N-1)v^4g^2}{8N}\big\langle A^\mu_{N^2-1}(x)A^\nu_{N^2-1}(y)\big\rangle_{\text{tl}} + 
\frac{v^2}{2}\partial^\mu_x\partial^\nu_y \big\langle h(x)h(y)\big\rangle_{\text{tl}} + \cdots\;.
\label{eq:fmsgauge}
\end{align}
The poles of the r.h.s. are at the mass $M_A$ of the heaviest gauge boson and at the mass $m_h$ of the 
elementary Higgs field. Depending on the details of the theory one of those masses will be the mass of the ground 
state of $O^\mu_{1^-_0}$. 

Interestingly, this implies a lower limit of one for the ratio of the scalar to the vector mass: Suppose $M_A > m_h$. 
This implies that the ground state mass of the vector singlet state is $m_{1^-_0} = m_h$ and thus the FMS 
mechanism predicts at tree level always $m_{1^-_0} \leq m_{0^+_0}$ 
\footnote{The same argument does not hold for $\SUtwo$ gauge theories because of the different custodial 
structure, but the same pattern is observed nonetheless in lattice calculations 
\cite{Maas:2014pba,Evertz:1985fc}.}. This is not the case in perturbation theory, where this ratio is free.

Let us now focus on states with open $\U$ quantum numbers \cite{Maas:2017xzh}. Since the corresponding 
charge is conserved, the lightest such state is absolutely stable. A scalar operator with open $\U$ quantum number 
is given by
\begin{align}
O_{0^+_1} &= \epsilon_{i_1\cdots i_n}\phi_{i_1}(D_{\mu_1}\phi)_{i_2}(F^{\mu_1}_{~\mu_2}\phi)_{i_3}
\cdots (F^{\mu_{N-3}}_{\quad\mu_{N-2}}\phi)_{i_{N-1}}(D^{\mu_{N-2}}\phi)_{i_N}\;,
\label{eq:u1scalar}
\end{align}
for $N>3$ and $\epsilon_{i_1i_2i_3}\phi_{i_1}(D_\mu\phi)_{i_2}(D_\nu F^{\mu\nu}\phi)_{i_3}$ for $N=3$. 
A vector state with open $\U$ quantum number is constructed in a similar way,
\begin{align}
O_{1^-_1}^\mu &= \epsilon_{i_1\cdots i_n}\phi_{i_1}(D_{\nu_1}\phi)_{i_2}(F^{\nu_1}_{~\nu_2}\phi)_{i_3}
\cdots (F^{\nu_{N-2}\mu}\phi)_{i_{N}}\;,
\label{eq:u1vector}
\end{align}
for $N>2$. Applying the FMS mechanism and employing a tree-level analysis to the bound state correlators of Eq. 
\eqref{eq:u1scalar} and \eqref{eq:u1vector}, see \cite{Maas:2017xzh}, yields for both states a ground state mass 
of $(N-1)m_A$ for any $N>2$. An analogous analysis can be done for the charged conjugated partners of both 
states and yields, of course, the same result. Since $\min(M_A,m_h)\leq 2m_A$ this implies that these states are 
always heavier than the singlet vector.

To briefly summarize the results of this section, we observe a qualitative difference between the elementary 
spectrum, which is usually used to describe the physical states of a gauge theory with a Brout-Englert-Higgs (BEH) 
effect, and the gauge-invariant observable spectrum. The two different spectra are recapped in Tab. \ref{tab:suf}. 
While the elementary spectrum contains only one scalar, a singlet with mass $m_h$, there are three physical scalar 
states. 
One of them is a singlet regarding the global $\U$ symmetry group, also with mass $m_h$. The other two are 
$\U$ non-singlets with mass $(N-1)m_A$ and describe the particle and anti-particle state of the open $\U$ 
channel. 
The spectrum also differs in the $1^-$ channel. The elementary spectrum contains $(N-1)^2-$ massless gauge 
boson, $2(N-1)$ mass-degenerate massive ones, and a single gauge boson with a heavier mass. The gauge-
invariant spectrum has two poles in the $\U$-singlet spectrum, one being a (potentially stable) second composite 
state, and two $\U$ non-singlet states, which are mass-degenerate with the scalar non-singlet and correspond 
again to particle and anti-particle.

Though the methods involved have a similar range of validity as standard perturbation theory, these results are 
quite extraordinary. They need therefore further confirmation. Such support has been provided using lattice 
calculations, as will be shown now.

\begin{table*}[t!]
\caption{Left: Elementary spectrum of an $\mathrm{SU}(N>2)$ gauge theory with a single scalar field in the fundamental 
representation. Right: Observable (physical) spectrum of the theory. Here $m_h$ denotes the mass of the 
elementary Higgs field, $M_A$ is the mass of the heaviest elementary gauge boson and $m_A$ the mass of the 
degenerated lighter massive gauge bosons. We assign a custodial $\U$ charge of $1/N$ to the scalar field $\phi$. 
The definition of the fields and operators can be found in the main text.}
\label{tab:suf}
\begin{center}
\begin{tabular}{c|ccc|cccc}

\multicolumn{4}{c}{elementary spectrum} & \multicolumn{4}{c}{gauge-invariant spectrum} \cr

\hline

\hline
$J^P$ & \; Field \; & \; Mass \; & \; Deg. \; & \; $\U$ \; & \; Operator \; & \; Mass \; & \; Deg. \; \cr
\hline
$0^+$ & $h$ & $m_h$ & $1$ & $0$ & $O_{0^+_0}$ & $m_h$ & $1$\cr
$\;$ & $\;$ & $\;$ & $\;$ & $\pm 1$ & $O_{0^+_{\pm 1}}$ & \quad $(N-1)m_A$ \; & $1/\bar{1}$\cr
\hline
$1^-$ & $A^\mu_{1,\dots,(N-1)^2-1}$ & $0$ & $(N-1)^2-1$ & $0$ & $O^\mu_{1^-_{0}}$ & $M_A$ or $m_{h}
$ & $1$\cr
 & $A^\mu_{(N-1)^2,\dots,N^2-2}$ & $m_A$ & $2(N-1)$ & $\pm 1$ & $O^\mu_{1^-_{\pm 1}}$ & $(N-1)m_A
 $  & $1/\bar{1}$ \cr
 & $A^\mu_{N^2-1}$ & $M_A$ & $1$  &  &  & & \cr
\hline

\hline
\end{tabular}

\end{center}
\end{table*}

\section{Lattice results for N=3}
The lattice action, employed for the creation of configurations, is the euclidean discretization of 
\eqref{eq:lagrangian},
\begin{align}
S[U,\phi] &= S_G[U] + \sum_x\Bigg[\phi(x)^\dagger\phi(x) + \lambda\Big(1-\phi(x)^\dagger\phi(x)\Big)^2 
-\kappa\sum_{\mu=\pm 1}^{\pm 4} \phi(x)^\dagger U_\mu(x)\phi(x+\hat{\mu}) \Bigg]\;,
\label{eq:action}
\end{align}
where the first term is the usual Wilson gauge action with inverse gauge coupling $\beta$ and link variables 
$U_\mu$ which describe the gauge bosons as 
$A_\mu = \frac{1}{2ai}(U_\mu - U_\mu^\dagger)|_{\text{traceless}}+{\mathcal O}(a^2)$.  The parameter 
$\lambda$ adjusts the four-Higgs coupling and $\kappa$ triggers the coupling between the Higgs field $\phi$ and 
the links $U_\mu$.

We used a standard multi-hit Metropolis algorithm to perform our analysis. We usually dropped $300+10L$ initial 
configurations for thermalization and $3L$ configurations in between measurements for decorrelation, where $L$ is 
the lattice size. 
We typically have $\mathcal{O}(10^4-10^5)$ configurations to measure gauge-invariant operators. For these 
operators we additionally STOUT-smeared the links and APE-smeared the Higgs field to enlarge our operator basis.

To test the FMS mechanism, we need a set of parameters of the action \eqref{eq:action} where a BEH effect 
takes place. To do this we scanned the phase diagram, using the methods of \cite{Caudy:2007sf}, to identify 
suitable parameter sets.

Further details on the simulations and results for the phase diagram can be found in 
\cite{Maas:2016ngo,Me2:unpublished}. Here, we will concentrate on the results at a sample point at 
$\beta = 6.85535$, $\kappa = 0.456074$, $\lambda = 2.3416$. This point is close to the boundary between the 
QCD-like region and the Higgs-like region where we expect the largest cutoffs, i.e. the smallest lattice spacings 
\cite{Maas:2014pba}. In fact, we find $am_{\text{l.p.}}=0.39$ with 'l.p.' denoting the lightest physical particle 
in the spectrum.

Since the propagators of elementary fields on the r.h.s. of the FMS expansion are all gauge-dependent, we need to 
fix a gauge.  We fix to 't Hooft-Landau gauge using stochastic overrelaxation to fix to Landau gauge first and 
then rotate by a global gauge transformation the Higgs expectation value in the real $3$-direction 
\cite{Maas:2016ngo}. This realizes the gauge choice of Subsection \ref{sec-2}, Eq. \eqref{eq:rxi}, in the limit
 $\xi\to 0$. We usually generate $\mathcal{O}(10^3-10^4)$ gauge-fixed configurations to measure the 
 propagators of the gauge-dependent elementary degrees of freedom.

We computed the propagators for the vector channel from the gauge fields on the gauge-fixed configurations as
\begin{align}
D^{bc}(p^2) &= \Big\langle A_\mu^b(-p)A_\mu^c(p)\Big\rangle\;.
\label{eq:schwinger}
\end{align}
Due to our choice of gauge, the propagator is diagonal in color space and thus 
$D^{bc}(p^2)=D^{b}(p^2)\delta^{bc}$. Further, $b=1,2,3$ correspond to the massless gauge bosons, 
$b=4,5,6,7$ correspond to the four degenerate massive gauge bosons and $b=8$ to the heaviest gauge boson. 
The degenerate gauge boson propagators are averaged over.  
If the propagators show tree-level behavior, then the effective masses can be obtained from the data by fitting to 
a simple tree-level propagator $Z/(p^2+a^2m_{\text{eff}}^2)$.

\begin{figure}[thb!] 
  \centering
  \includegraphics[width=0.482\textwidth,clip]{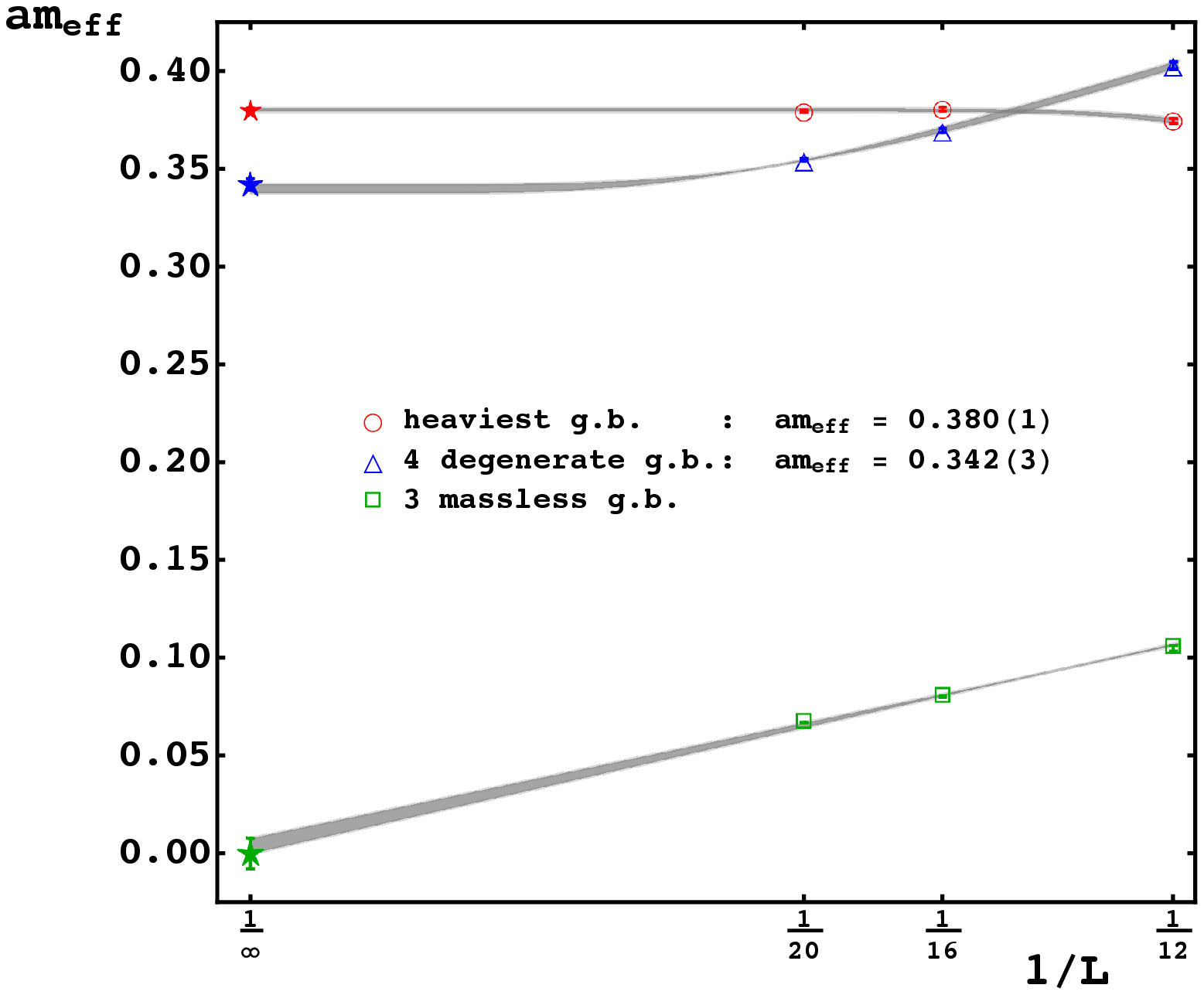}
   \includegraphics[width=0.495\textwidth,clip]{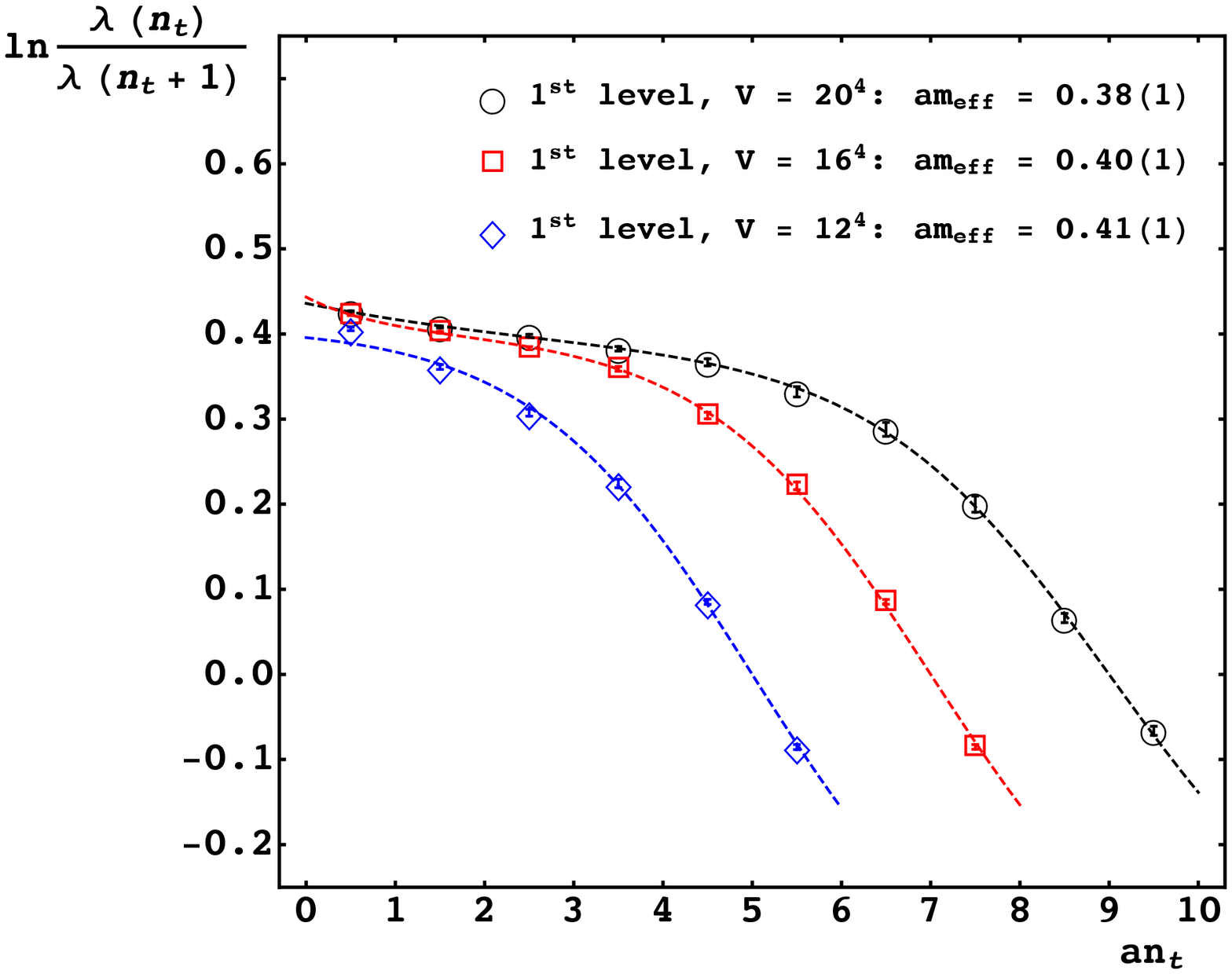}
  \caption{Left: Effective masses from the (gauge-variant) propagators for the four degenerate massive 
  (blue triangles), the heaviest (red circles) and the three massless gauge bosons (green boxes) as a 
  function of the inverse lattice size $L$. The gray bands are error bands obtained from fits of the data to 
  $am_{\text{eff}}+Be^{-cV}$ for the massive modes and to $C+D/L$ for the massless modes.
  Right: Ground state of the (gauge-invariant) vector $\U$-singlet channel obtained from a variational  
  analysis on $V=12^4$, $16^4$ and $20^4$ lattices. 
  The dashed lines are obtained from fits to the correlators of the form $\lambda(n_t) = 
  A_1\cosh\big(am_{\text{eff}}(n_t-L/2)\big)+A_2\cosh\big(am_{\text{eff}}^{\prime}(n_t-L/2)\big)$.}
  \label{fig:schwinger}
\end{figure}

Since the propagators behave perturbatively, which is confirmed by a measurement of the running gauge coupling 
to be presented elsewhere \cite{Me2:unpublished}, we could extract the effective mass in the described way. This 
was done for the lattice volumes $V=12^4$, $16^4$ and $20^4$. 
The results are shown in the left panel of Fig. \ref{fig:schwinger}. The gray bands are error bands obtained by fits 
of the masses to $am_{\text{eff}}+Be^{-cV}$ for the perturbatively massive modes ($b=4,5,\dots,7$, blue 
triangles and $b=8$, red circles) and to $C+D/L$ for the perturbatively massless modes ($b=1,2,3$, green boxes). 
The lattice masses extracted from the fits for the four degenerate gauge bosons are $am_A = 0.34(1)$ and for 
the heaviest gauge boson is $aM_A=0.38(1)$, for $L\to \infty$. The ratio of those masses is $0.89(5)$ which is 
in good agreement of tree-level perturbation theory, where $m_A/M_A = \sqrt{3/4} \approx 0.87$ (see Eq. 
\eqref{eq:gaugemass}). 

At this point we want to mention that also the position-space propagators were computed. The 
masses extracted from those are in agreement, within considerably larger errors, with the ones above.

The right panel of Fig. \ref{fig:schwinger} shows the result of a variational analysis in the vector $\U$-singlet 
channel. We used in this channel three operators in multiple smearing levels to perform this analysis. The base 
operators read
\begin{align}
\begin{split}
O_j^{\phi U\phi}(t) &= \sum_{\bf x}\phi({\bf x},t)^\dagger D_j \phi({\bf x},t) 
\quad,\quad 
O_j^{\phi^3 U\phi}(t) = \left[ \sum_{\bf x}\phi({\bf x},t)^\dagger\phi({\bf x},t)\right]O_j^{\phi U\phi}(t)
\quad, \\ 
O_j^{(\phi U\phi)^3}(t) &= \sum_{k=1}^3\left[O_k^{\phi U\phi}(t)O_k^{\phi U\phi}(t)\right]O_j^{\phi U\phi}(t)
\;.
\end{split}
\label{eq:interpol}
\end{align}
These were smeared three and four times yielding a total set of six operators.
Details on this will be presented elsewhere \cite{Me2:unpublished}. We only show the lowest levels (ground states), 
for $V=12^4$, $V=16^4$ and $V=20^4$ lattices. The masses are extracted from double-$\cosh$ fits of the 
correlators (dashed lines). 
The extrapolated lattice mass ($L\to\infty$) of this state is $am_{1^-_0}=0.39(1)$ which is in good agreement 
with the mass extracted from the heaviest gauge boson propagator. Higher levels are yet too noisy and need more 
statistics for definite results. 
However, they are indeed high up in the spectrum, and compatible with the location of the elastic threshold at 
$3m_{1^-_0}$. For a more detailed analysis, including finite volume studies, see 
\cite{Maas:2016ngo,Me2:unpublished}. 

Thus, our result is that in the vector $\U$-singlet channel only one state appears below the elastic threshold, 
having the same mass as the heaviest gauge boson. This fully supports the prediction of the FMS mechanism,  
Eq. \eqref{eq:fmsgauge}.

The results for the $0^+_0$ channel will be presented elsewhere \cite{Me2:unpublished}, but we do not see a 
signal below the elastic threshold at $2m_{1^-_0}$. This is also in agreement with the prediction of $m_h>m_A$ 
if $m_{1^-_0}\approx m_A$. Unfortunately, this precludes us for now from checking the prediction in the scalar 
singlet channel without first searching for unstable resonances. Note that we cannot use the gauge-dependent 
propagator alone to predict $m_h$ because of the need for renormalization \cite{Maas:2013aia}.

\begin{figure}[t!]
  \centering
  \includegraphics[width=0.47\textwidth,clip]{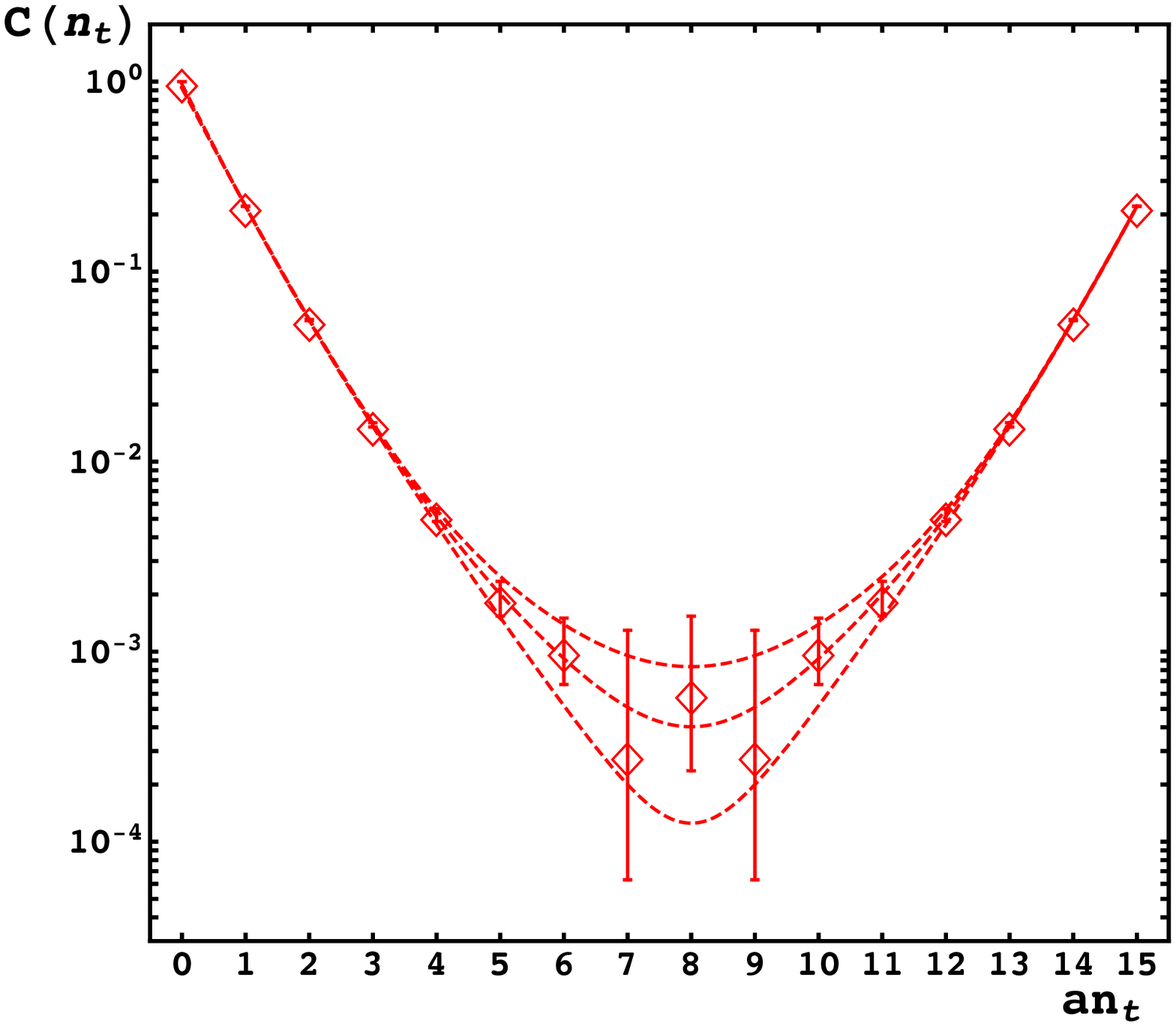}
  \includegraphics[width=0.515\textwidth,clip]{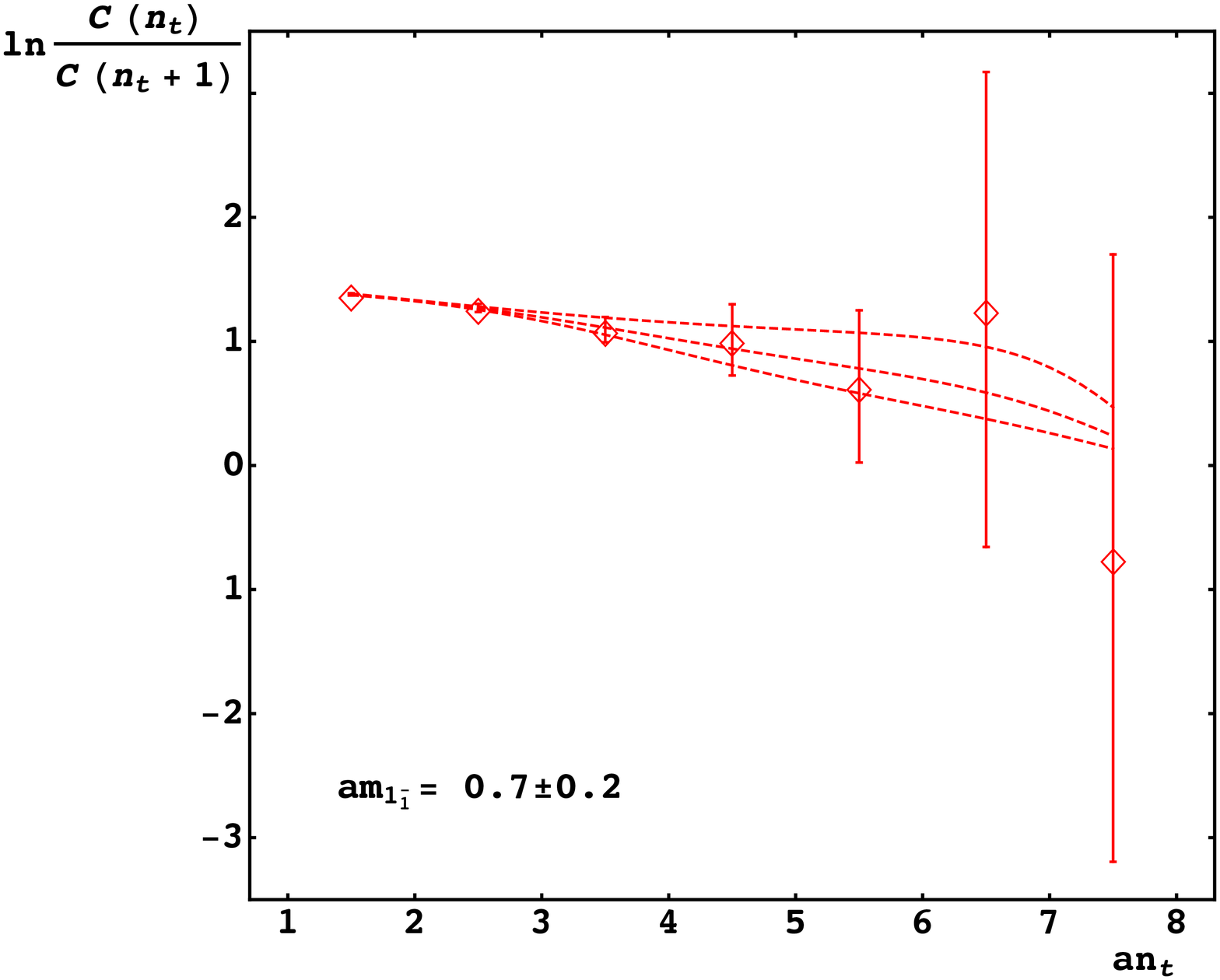}
  \caption{Left: Correlator of the open $\U$ state in the vector channel for a $V=16^4$ lattice. 
  Right: Effective mass extracted from the correlator. The dashed lines are obtained from fits to the correlator of 
  the form $C(n_t) = A_1\cosh\big(am_{\text{eff}}(n_t-L/2)\big)
  +A_2\cosh\big(am_{\text{eff}}^{\prime}(n_t-L/2)\big)$.}
  \label{fig:u1open}
\end{figure}

Let us now focus on the even more non-trivial predictions for the $\U$-non-singlet states. Since our statistics are 
yet insufficient for the scalar channel, we focus here on the vector channel. Again, more details will be available in 
\cite{Me2:unpublished}. In Fig. \ref{fig:u1open} the correlator (left) and the corresponding effective mass (right) 
are shown. 

At this point we want to mention that slightly different operators with the same quantum numbers were used in 
contrast to Eq. \eqref{eq:u1scalar} and \eqref{eq:u1vector}, 
\begin{align}
O_{0^+_1} &= \epsilon_{ijk}\phi_i (D_\mu\phi)_j (D_\mu D^2\phi)_k \quad\text{and}\quad
O_{1^-_1}^\mu =  \epsilon_{ijk}\phi_i (D^\mu\phi)_j (D^2\phi)_k\;.
\label{eq:u1sv2}
\end{align}
The only reason why we used these operators is, that they are easier to implement. Both operators expand also to 
a ground state mass of $2m_A$, i.e. two times the mass of the lightest gauge boson. This is exactly the same 
mass as predicted for the two operators presented in Subsection \ref{sec-3}. However, these operators are special 
to the $N=3$ case, and cannot be trivially generalized to $N>3$, in contrast to Eq. \eqref{eq:u1scalar} and
\eqref{eq:u1vector}.

Using analogous methods as before we extract an effective mass of the open $\U$ vector state as 
$am_{1^-_1}=0.7(2)$ which is in agreement with the FMS prediction, i.e. $2am_A = 0.68(1)$. Of course higher 
statistics is needed to make a definite statement. It is, however, substantially lower than a naive expectation of 
$3am_h\ge6aM_A>2.3(1)$, which would be obtained in a naive quantum-mechanical counting of the mass of the 
operator in Eq. \eqref{eq:u1sv2}.

\section{Conclusions}\label{sec-4}

Summarizing, we used the FMS mechanism to predict the physical, experimentally observable, and gauge-invariant 
mass spectrum of an SU($N>2$) gauge theory with a single Higgs field in the fundamental representation in the 
custodial singlet and non-singlet scalar and vector channels. The results are qualitatively different from the 
perturbative predictions, which yields the spectrum of the elementary fields. These two predictions are compared in 
Tab. \ref{tab:suf}. 

We checked these predictions in the case of $N=3$ using non-perturbative lattice simulations. We determined the 
gauge-invariant and gauge-dependent spectrum. Our results support the predictions of the FMS mechanism, 
including non-trivial relations between the masses in different channels. We also showed that the theory is 
nonetheless weakly-coupled.

Of course larger volumes, larger parts of the parameter space, and higher statistics is needed to improve the 
support in the non-singlet channels. However, the qualitative features of the spectrum are already in agreement 
with the predictions, and this makes it somewhat unlikely that quantitative corrections will be able to alter the 
result substantially.

Thus it is possible to predict analytically, entirely on basis of gauge invariance and the FMS mechanism, the correct 
mass spectrum. Thus, in addition to the case of SU(2) with a single fundamental Higgs field, where the prediction of 
a coinciding spectrum of physical and elementary degrees of freedom was confirmed in lattice simulations 
\cite{Maas:2012tj,Maas:2013aia}, the results here constitute a highly non-trivial test of the underlying 
field-theoretical concepts. 
               
Besides improving the present results, tests could be made for other theories. Corresponding predictions have 
been made for a wide range of similar theories in \cite{Maas:2017xzh}. Eventually, this can, and should, be 
extended to current candidates for BSM physics, as has already been done for the full standard model 
\cite{Frohlich:1980gj,Frohlich:1981yi,Egger:2017tkd}. The results here suggest that also for more realistic theories 
conflicts may easily arise.

\begin{acknowledgement}
\subsection*{Acknowledgements} 

PT has been supported by the FWF doctoral school W1203-N16. The computational results 
presented have been obtained using the Vienna Scientific Cluster (VSC) and the HPC center 
at the University of Graz. RS thanks the Carl-Zeiss Stiftung for support through a postdoc 
fellowship.
\end{acknowledgement}

\bibliography{lattice2017}

\end{document}